\documentclass[sigconf]{acmart}
\AtBeginDocument{%
  }

\usepackage{enumitem}
\usepackage{multirow}
\usepackage{makecell}
\usepackage{subcaption}

\usepackage{amsmath}
\usepackage{tabularx}
\setlist[itemize]{left=0pt}
\copyrightyear{2026}
\acmYear{2026}
\setcopyright{cc}
\setcctype{by}
\acmConference[CIKM '26]{Proceedings of the 35th ACM International Conference on Information and Knowledge Management}{November 07--11, 2026}{Rome, Italy}
\acmBooktitle{Proceedings of the 35th ACM International Conference on Information and Knowledge Management (CIKM '26), November 07--11, 2026, Rome, Italy}
\acmDOI{10.1145/3799682.3841034}
\acmISBN{979-8-4007-2539-5/2026/11}
\begin{document}

\title{Can LLM Annotations Replace User Clicks for Learning to Rank?}

\author{Lulu Yu}
\affiliation{%
\institution{State Key Laboratory of AI Safety}
\institution{Institute of Computing Technology, Chinese Academy of Sciences}
\institution{University of Chinese Academy of Sciences}
  \city{Beijing}
  \country{China}}
\email{nothing_0_1@163.com}

\author{Keping Bi}
\authornote{Corresponding author.}
\affiliation{%
\institution{State Key Laboratory of AI Safety}
\institution{Institute of Computing Technology, Chinese Academy of Sciences}
\institution{University of Chinese Academy of Sciences}
  \city{Beijing}
  \country{China}}
\email{bikeping@ict.ac.cn}

\author{Jiafeng Guo}
\affiliation{%
\institution{State Key Laboratory of AI Safety}
\institution{Institute of Computing Technology, Chinese Academy of Sciences}
\institution{University of Chinese Academy of Sciences}
  \city{Beijing}
  \country{China}}
\email{guojiafeng@ict.ac.cn}

\author{Shihao Liu}
\author{Shuaiqiang Wang}
\author{Dawei Yin}
\affiliation{
\institution{Baidu Inc.}
\city{Beijing}
\country{China}
}
\email{liushihao@baidu.com}
\email{wangshuaiqiang@baidu.com}
\email{yindawei@acm.org}

\author{Xueqi Cheng}
\affiliation{%
\institution{State Key Laboratory of AI Safety}
\institution{Institute of Computing Technology, Chinese Academy of Sciences}
\institution{University of Chinese Academy of Sciences}
  \city{Beijing}
  \country{China}}
\email{cxq@ict.ac.cn}

\renewcommand{\shortauthors}{Lulu Yu et al.}

\begin{abstract}
Large-scale supervision is essential for training modern ranking models, yet obtaining high-quality human relevance annotations remains costly. Click data has long been used as a low-cost alternative, and recent advances in large language models (LLMs) have introduced LLM-based relevance annotation as another promising source of supervision. This paper investigates whether LLM annotations can replace click data for learning to rank (LTR) through a comprehensive comparison across multiple dimensions. Experiments on both a public dataset, TianGong-ST, and an industrial dataset, Baidu-Click, show that click-supervised models perform better on high-frequency queries, whereas LLM-annotation-supervised models are more effective on middle- and low-frequency queries. Further analysis suggests that click-supervised models better capture document-level signals, such as authority and quality, while LLM-annotation-supervised models are more effective at modeling semantic matching between queries and documents and distinguishing relevant from non-relevant documents. Motivated by these findings, we explore two training strategies, annotation-aware data scheduling and frequency-aware multi-objective learning, to integrate both supervision signals. Both strategies improve ranking performance across queries at all frequency levels, with frequency-aware multi-objective learning achieving stronger results. Our code is available at \url{https://github.com/Trustworthy-Information-Access/LLMAnn_Click}.
\end{abstract}

\ccsdesc[500]{Information systems~Retrieval models and ranking}

\keywords{Unbiased Learning to Rank, Large Language Models, Relevance Annotation, Learning to Rank}



\maketitle

\section{Introduction}

Learning to rank (LTR) plays a central role in modern search engines and other information access systems. Its goal is to train ranking models that order documents according to their relevance to a given query. Early LTR methods are typically based on tree-based models such as LambdaMART \cite{burges2010ranknet}, relying on hand-crafted query-document features such as BM25 \cite{robertson1994some}. With the rapid development of deep learning, especially Transformer architectures \cite{vaswani2017attention}, modern ranking models are increasingly implemented as cross-encoders \cite{alaparthi2020bidirectional,yates2021pretrained}, which jointly encode queries and documents to capture fine-grained semantic interactions. However, these models contain a large number of parameters and therefore require substantial supervision for effective training. Since large-scale human relevance annotations are expensive to obtain, low-cost supervision signals have become essential for training practical ranking systems.

Search logs provide abundant user interaction signals, among which clicks are the most widely used. Clicks can be collected at scale and serve as implicit indicators of query-document relevance. However, click signals are inherently biased \cite{gupta2024unbiased}, affected by factors such as position bias \cite{ai2021unbiased,joachims2017unbiased,wang2018position}, selection bias \cite{agarwal2019addressing,vardasbi2020inverse,vardasbi2021mixture}, and trust bias \cite{ovaisi2020correcting,wang2016learning,zhao2023unbiased}. This has motivated extensive research on unbiased learning to rank (ULTR), which aims to correct such biases and learn unbiased ranking models. Due to the limited availability of large-scale real-world click datasets, many ULTR methods are evaluated on simulated clicks generated from assumed user browsing models \cite{richardson2007predicting,vardasbi2020cascade}. In practice, however, real-world clicks reflect more complex and nuanced user behavior \cite{hager2024unbiased,yu2025unbiased,zou2022large}. As a result, effectively leveraging and debiasing real-world click data remains a challenging problem.

Recently, large language models (LLMs) have shown strong potential as relevance annotators \cite{dewan2025llm,rahmani2025judging,thomas2024large,zhang2024large}. Although LLM-based annotation incurs inference costs, it provides a scalable alternative to expensive human judgments. Compared with clicks, LLM annotations are less affected by user behavior biases and can provide more flexible relevance judgments. This naturally raises an important question: can LLM annotations replace click data for training ranking models, or do the two supervision sources capture different relevance signals?

To answer this question, we first compare the annotation characteristics of click data and LLM annotations, as illustrated in Fig.~\ref{fig:comparison}. We identify two major differences. First, LLMs offer greater flexibility in generating annotations, since they can produce pointwise, pairwise, or listwise relevance judgments under different annotation guidelines. Second, the relevance criteria reflected by the two sources are different. Click data can capture not only semantic matching between queries and documents, but also document-level signals such as authority, quality, and attractiveness. In contrast, LLM annotations are effective at judging semantic relevance between query-document pairs, but may be less sensitive to document-level signals that influence user clicks.
{\setlength{\belowcaptionskip}{-7mm}
\begin{figure}
    \centering
    \includegraphics[width=1.0\linewidth]{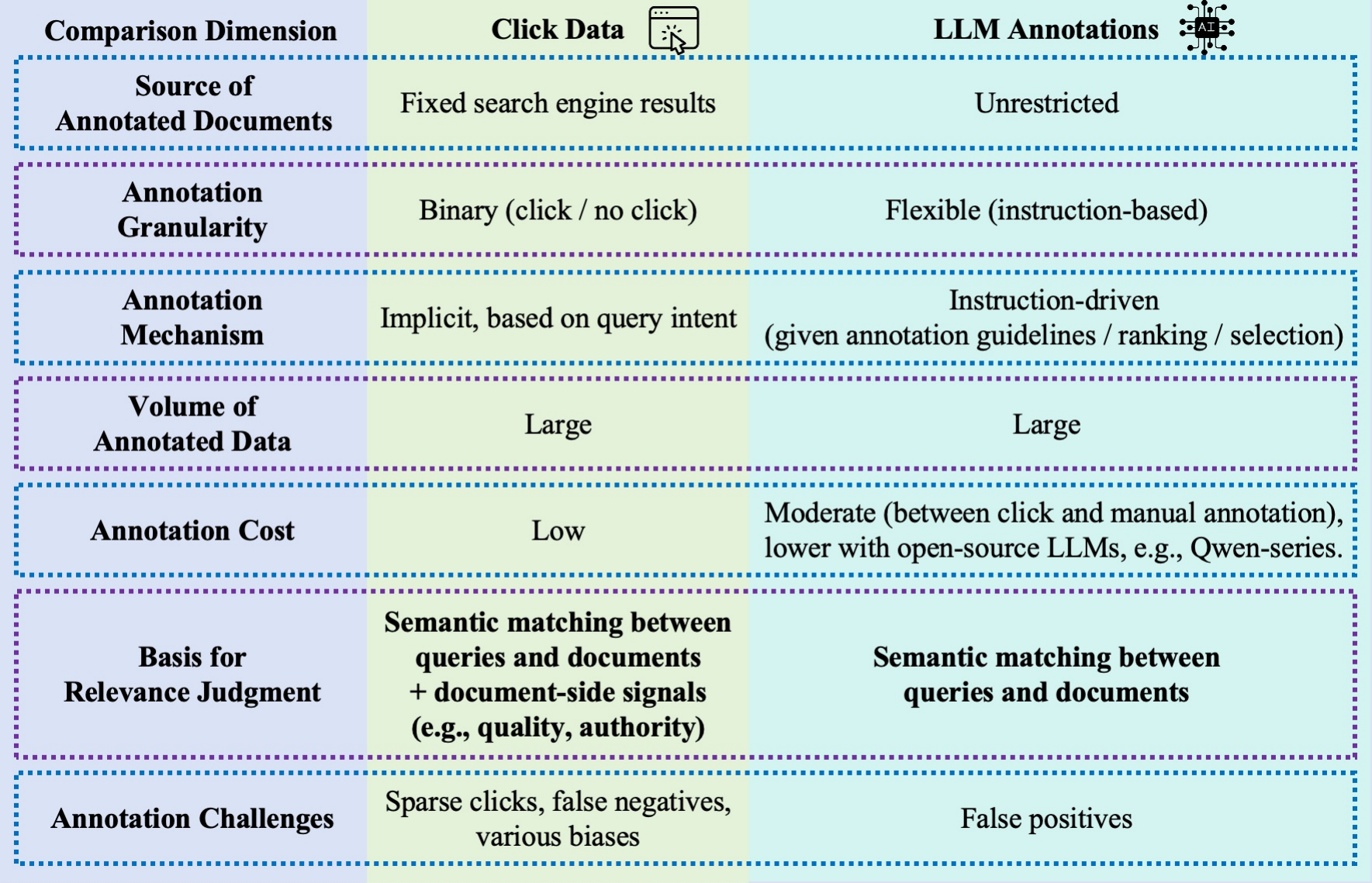}
    \vspace{-6mm}
    \caption{Comparison of annotation characteristics between click data and LLM annotations.}
    \label{fig:comparison}
\end{figure}}

We then design three LLM-based annotation strategies: generating multi-level relevance labels according to annotation guidelines, directly ranking a list of candidate documents, and directly selecting relevant documents. Our experiments show that, under the given annotation guidelines, listwise annotation with a query and a list of documents leads to the best ranking performance.

Beyond comparing annotation strategies, we further examine how ranking models trained with clicks and LLM annotations behave across queries with different frequencies. Since search engines tend to perform better on high-frequency queries by returning more relevant results \cite{zou2022large}, models trained on high-frequency click data may benefit from richer document-level signals. We therefore compare click-trained and LLM-annotation-trained models across different query-frequency groups. To better understand their relevance modeling capabilities, we also analyze the performance gains obtained by incorporating different groups of LTR features, including document-level features such as PageRank and URL structure, and query-document matching features such as BM25 \cite{ai2018unbiased,chen2023thuir,robertson1994some,zhai2017study}. The results show that click-trained models perform better on high-frequency queries, whereas models trained with LLM annotations have advantages on middle- and low-frequency queries. Our feature-level analysis further suggests that click-trained models are more effective at exploiting document-level signals, while LLM-annotation-trained models better capture semantic matching between queries and documents.

We also investigate whether ranking models trained with these two types of supervision can distinguish relevant from irrelevant documents. Since the training set contains a high proportion of relevant results, it is unclear whether the models truly learn relevance discrimination or mainly learn to rank among already relevant documents. To study this issue, we introduce true negative samples into training and evaluation. The results show that models trained with LLM annotations have a stronger ability to distinguish relevant from irrelevant documents than models trained with clicks. For click-trained models, adding true negative samples improves both their ability to separate relevant from irrelevant documents and their ability to distinguish different degrees of relevance among relevant documents. However, for models trained with LLM annotations, true negative samples can be detrimental, as they encourage a stronger separation between relevant and irrelevant documents and weaken the model's ability to capture fine-grained relevance differences.

These findings suggest that click data and LLM annotations provide complementary supervision signals. Click data is especially useful for high-frequency queries and for capturing document-level signals, while LLM annotations are more effective for middle- and low-frequency queries and for modeling semantic relevance. Based on this observation, we propose two methods to integrate the two sources of supervision. The first is annotation-aware data scheduling. Since LLM annotations provide strong semantic supervision, we first warm-start the ranking model using LLM-annotated data. We then perform interleaved training with click data from high-frequency queries and LLM annotations from middle- and low-frequency queries to prevent forgetting and exploit the strengths of both sources. The second is frequency-aware multi-objective learning, which uses query frequency as a gating signal to adaptively adjust the weights of click-based and LLM-based objectives during joint training. Both methods improve ranking performance by leveraging the complementary advantages of click data and LLM annotations, with frequency-aware multi-objective learning achieving stronger results.

In summary, our contributions are as follows:
\begin{itemize}
    \item We conduct a comprehensive empirical comparison of ranking models trained with click data and LLM annotations from multiple perspectives, including annotation strategy, query frequency, feature-level relevance signals, and relevance discrimination ability.
    \item We show that click-trained models are more effective at exploiting document-level signals and perform better on high-frequency queries, while LLM-annotation-trained models better capture semantic matching, perform better on middle- and low-frequency queries, and more effectively distinguish relevant from irrelevant documents.
    \item We propose two hybrid training strategies that integrate click data and LLM annotations, demonstrating that their complementary strengths can be effectively combined to improve ranking performance.
\end{itemize}
\section{Related Work}
This section reviews related work on learning to rank with click data and recent advances in using large language models (LLMs) for ranking, including LLM-based relevance annotation and zero-shot LLM ranking.

\subsection{Learning to Rank with Click Data}

As ranking models become increasingly data-intensive, learning to rank (LTR) requires large amounts of supervision. Since large-scale human relevance annotations are expensive to obtain, click data has been widely used as a low-cost alternative. Clicks provide implicit feedback from real users and can be collected at scale, making them an important source of supervision for industrial search systems. However, clicks are also noisy and biased, which has motivated extensive research on unbiased learning to rank (ULTR) \cite{ai2018unbiased,chen2021adapting,luo2023unconfounded,zhang2023towards}.

Existing ULTR methods mainly fall into two categories: click modeling and counterfactual learning. Click modeling approaches \cite{chuklin2022click,craswell2008experimental} explicitly model user browsing behavior to infer relevance from observed clicks. These methods often require repeated observations of the same query--document pairs, which limits their effectiveness on sparse or long-tail queries. Counterfactual learning methods instead use inverse propensity weighting (IPW) to correct click bias \cite{joachims2017unbiased,wang2018position}. Since the effectiveness of IPW depends heavily on accurate propensity estimation, several studies jointly estimate ranking and propensity models \cite{ai2018unbiased}.

Early ULTR methods mainly focus on position bias, assuming that examination propensity depends only on rank. In practice, however, click behavior is affected by more complex factors, including surrounding results \cite{chen2021adapting}, selection effects, trust bias, and the strength of the production ranker. A strong ranker can induce relevance--position coupling, which leads to propensity overestimation and severe false negatives. Recent work addresses these issues through relevance--position decoupling \cite{zhang2023towards}, unconfounded propensity estimation \cite{luo2023unconfounded}, and revised click hypotheses such as DualIPW \cite{yu2025unbiased}.

Despite these challenges, click data remains valuable for LTR because it reflects real user interactions and captures signals beyond semantic relevance, such as document authority, quality, attractiveness, and practical utility. These signals are especially useful for high-frequency queries, where many results may be semantically relevant and clicks help distinguish which documents users actually prefer.
However, click data is also incomplete as supervision. It is usually limited to top-ranked results examined by users, leading to few truly irrelevant examples and weakening relevance discrimination. Reliable click signals also require sufficient observations, making them less effective for middle- and low-frequency queries. Moreover, existing debiasing methods often rely on simplified assumptions that may not fully capture real-world user behavior.

\subsection{LLM-Based Relevance Annotation}

Large language models (LLMs) have demonstrated strong language understanding and reasoning abilities \cite{mann2020language,team2024qwen2}, which has led to growing interest in using them for automatic annotation \cite{gilardi2023chatgpt,huang2023chatgpt,kuzman2023chatgpt}. In information retrieval, LLMs have been used to generate relevance judgments, construct evaluation datasets, and assess document utility. Prior work shows that LLM-generated relevance judgments can be comparable to crowd-sourced annotations \cite{thomas2024large}. Other studies further distinguish relevance from utility and show that utility-oriented annotations can improve retrieval-augmented generation (RAG) \cite{zhang2024large}.

Most existing work on LLM-based annotation focuses on dataset construction or evaluation \cite{dewan2025llm}, while less attention has been paid to how effective LLM annotations are as training supervision for ranking models. Some studies compare LLM-generated training data with conventional supervision signals \cite{askari2023test,bonifacio2022inpars,dai2022promptagator}. For example, \citet{zhang2025leveraging} show that models trained with LLM annotations can outperform models trained with human annotations in out-of-domain retrieval and RAG settings. However, comprehensive comparisons between LLM annotations and other low-cost supervision signals, especially click data, remain limited for ranking tasks.

LLM annotations have several appealing properties. They can directly assess semantic relevance between queries and documents, are not affected by user examination behavior, and can be flexibly generated under different annotation formats, such as pointwise, pairwise, or listwise judgments. These properties make LLM annotations especially promising for middle- and low-frequency queries, where click signals are sparse or unreliable. Moreover, because LLMs can judge documents that were not frequently exposed to users, they may provide better coverage than click data for training ranking models.

At the same time, LLM annotations have their own limitations. Recent work suggests that fairly wide confidence intervals may be needed to align LLM annotations with human annotations, indicating discrepancies between the two sources \cite{oosterhuis2024reliable}. LLMs can also exhibit position bias when processing multiple documents in a long context, particularly for documents placed in the middle \cite{liu2024lost}. In addition, annotation quality is sensitive to prompt design and annotation criteria \cite{alaofi2024llms}. More importantly for LTR, LLMs may be less effective at capturing document-level signals such as authority, quality, popularity, or other utility-related factors that are not fully expressed in the text. These signals can be crucial for high-frequency queries, where candidate documents may be semantically similar but differ substantially in user-perceived usefulness. Therefore, while LLM annotations provide strong semantic supervision, they may not fully replace the user-centric signals encoded in click data.

\subsection{Zero-shot Ranking with LLMs}

Another related line of work directly uses LLMs as rankers. In re-ranking scenarios, the candidate set is usually small enough for multiple documents to be provided to an LLM, allowing the model to generate rankings directly \cite{ma2023zero,sun2023chatgpt,zhuang2024setwise}. This approach benefits from the strong instruction-following and semantic reasoning abilities of large LLMs, but it also requires the LLM itself to possess strong ranking capability. Smaller LLMs often struggle in this setting, so prior work has explored distilling rankings generated by powerful LLMs into smaller models \cite{pradeep2023rankvicuna,pradeep2023rankzephyr,sun2023chatgpt}.

However, direct LLM ranking is costly for practical deployment. Generating a full ranking often requires feeding multiple documents into the LLM and performing several passes due to context length constraints. This leads to substantial inference overhead, especially in large-scale search systems. In contrast, our work focuses on using LLMs as annotation sources rather than online rankers. By distilling LLM-generated relevance judgments into conventional ranking models, we aim to obtain lightweight ranking models that benefit from LLM supervision without incurring the high online cost of LLM-based re-ranking.

\section{LLM Annotations and Analysis}
Given the high cost of human annotations, it is typically conducted by labeling each query-document pair individually according to pre-defined guidelines. Meanwhile, click signals generally provide only single-level supervision. In contrast, LLMs enable a highly flexible approach to relevance annotation. In this section, we describe how relevance annotations are generated using LLMs and present a relatively detailed analysis of the annotations.
\subsection{Annotation Strategy}
The flexibility of LLM-based relevance annotation is reflected in two main aspects: 1) the input format of query-document pairs, which can vary depending on the prompt design; and 2) the manner in which annotations are generated—LLMs can produce multi-level relevance annotations according to different annotation guidelines, or directly perform relevance selection or ranking over a list of candidate documents.
\begin{figure*}[!t]
\centering
\includegraphics[width=0.95\linewidth]{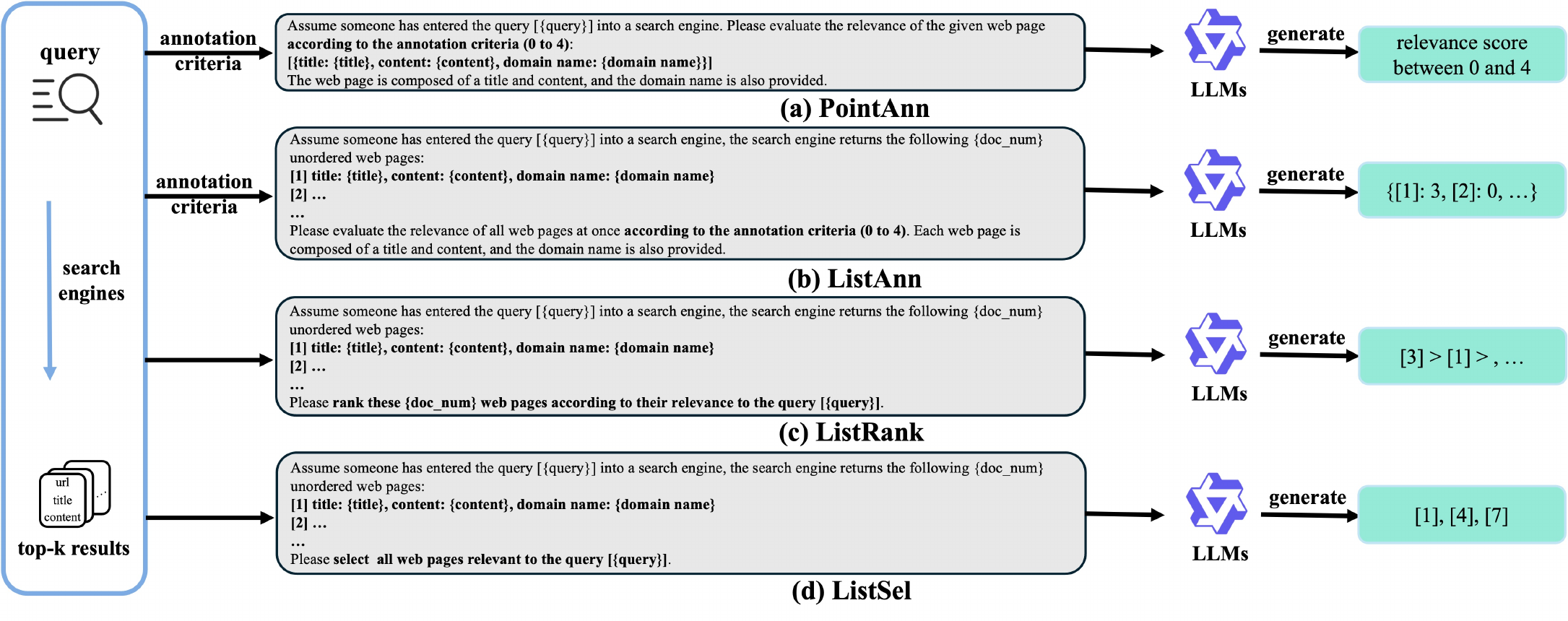}
\vspace{-6mm}
\caption{Annotation strategies with LLMs: (a) PointAnn, (b) ListAnn, (c) ListRank, and (d) ListSel.}
\vspace{-2mm}
\label{fig:llm_ann}
\end{figure*}
As shown in Fig. \ref{fig:llm_ann}, we employ three primary approaches to perform relevance annotation using LLMs:
\begin{itemize}
\item Multi-level annotation based on human-defined guidelines: LLMs generate graded relevance annotations following following the same human-defined relevance criteria used for human annotations of the corresponding dataset. This approach can be further divided into two types based on the input format: \textbf{PointAnn}, single query-document pair input; and \textbf{ListAnn}, single query with multiple candidate documents as input.
\item Listwise Ranking (\textbf{ListRank}): The LLM is prompted to directly rank a list of candidate documents based on their relevance to the query.
\item Listwise Selection (\textbf{ListSel}): The LLM is tasked with selecting a subset of documents from a given list that are relevant to the query.
\end{itemize}

Note that the document lists are input in the order they appeared in the original click logs. We assume that the original order reflects a strong underlying ranking model, and when the input order exhibits higher ranking quality, the LLM is more likely to generate higher-quality annotations \cite{zhuang2024setwise}.

LLM annotations may depend on prompt design and annotation criteria. In this work, we use standard IR relevance assessment guidelines with a 5-level scale (0--4), where label 2 denotes basic relevance and labels 3--4 additionally consider factors such as timeliness and authority. We leave different criteria or prompting strategies, such as pairwise prompting \cite{qin2024large}, for future work.

Each of the three annotation strategies has its own strengths and weaknesses in terms of annotation accuracy, cost, and effectiveness in training ranking models. We utilize 8 $\times$ A800 GPU with a cost of \$0.8 per hour for a single A100 GPU\footnote{\url{https://vast.ai/pricing/gpu/A800-PCIE}}, and employ Qwen2.5-32B-Instruct \cite{team2024qwen2} for annotation. Annotations with more LLM backbones are left for future work. We annotate approximately 2.6 million query–document pairs, and the annotation time for PointAnn, ListAnn, ListRank, and ListSel is 52.0, 7.6, 6.9, and 6.1 hours, respectively. Since we annotate the top ($\leq 10$) results returned from search logs, all documents can be input to the LLMs at once. Although listwise input involves more content and generates more output, the overall annotation time remains short. Multi-level annotation provides fine-grained supervision but can be highly sensitive to the given annotation criteria. Ranking-based and selection-based annotations rely on comparisons among multiple documents, which are constrained by the requirement to present multiple documents at once for comparison. This is particularly problematic for ranking-based methods, where obtaining a global ranking requires multiple annotations. In addition, meaningful permutations depend on the LLM’s ability to rank effectively. Nonetheless, it offers more fine-grained signals of relative relevance between documents. Effectively leveraging such signals is critical. In contrast, selection-based annotations provide binary annotation signals, which may lack sufficient granularity to serve as strong supervision when used for model training.


\subsection{Consistency with Human Annotations}
Furthermore, we analyze the alignment between different annotation methods and human annotations. Considering that the ultimate goal is to train ranking models using annotated data, the relative relevance ranking within a document list is more important than individual annotations. Therefore, we primarily focus on the consistency of ranking orders between multi-level annotations and human-annotated rankings, as well as their ranking performance. We employ the Spearman correlation \cite{hauke2011comparison} and the Kendall correlation \cite{abdi2007kendall} to measure ranking consistency, and nDCG@10 \cite{jarvelin2002cumulated} on the test set to evaluate ranking effectiveness.

\begin{figure}[!t]
    \centering
    \begin{subfigure}[b]{0.24\textwidth}
      \includegraphics[width=\textwidth]{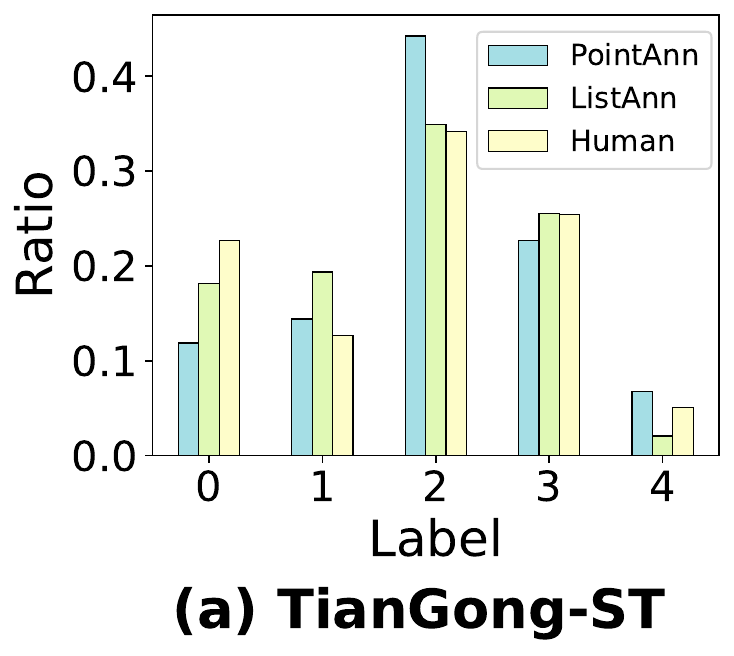}
      \label{fig:single_perf}
    \end{subfigure}%
    ~
    \begin{subfigure}[b]{0.24\textwidth}
      \includegraphics[width=\textwidth]{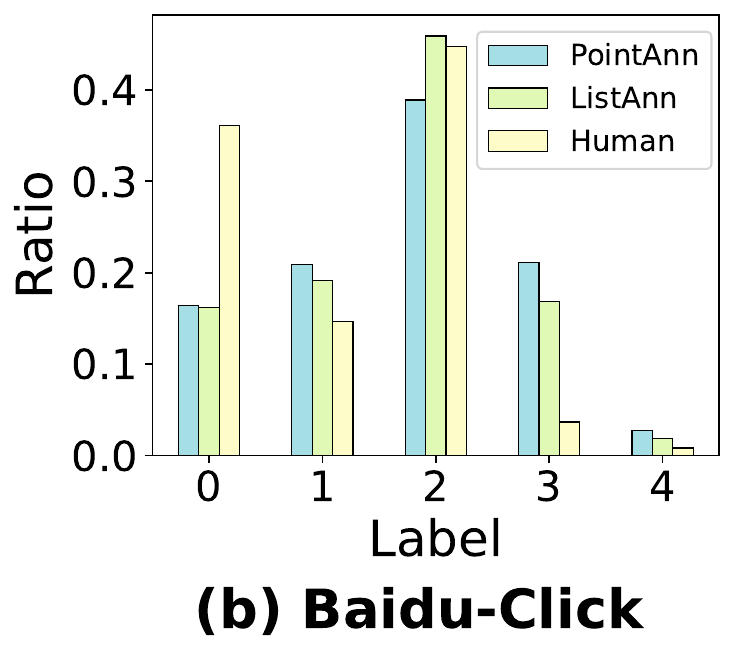}
      \label{fig:single_dis}
    \end{subfigure}
    \vspace{-12mm}
    \caption{Distribution of PointAnn, ListAnn, and human annotations at various relevance levels.}
    \label{fig:ann_dist_comp}
    \vspace{2mm}
\end{figure}

\enlargethispage{1\baselineskip}
Before conducting the consistency analysis, we first examine the distribution of LLM annotations, including PointAnn and ListAnn, as their formats are completely consistent with human annotations. This allows us to more directly perceive the characteristics of LLM annotations. As shown in Fig. \ref{fig:ann_dist_comp}, compared to human annotations, the proportion at label 0 of LLM annotations drops sharply, suggesting that LLMs impose a looser relevance threshold during annotation, which increases the risk of false positives. Here, label 2 corresponds to basic relevance, while labels 3 and 4 further consider timeliness and authority. On the Baidu-Click dataset, LLM annotations tend to more aggressively assign documents to labels 3 and 4, possibly because the returned results are generally of higher quality and LLMs inherently have limited ability to capture document-level signals, making them more easily misled.
\setlength{\belowcaptionskip}{-6mm}
Considering that consistency metrics such as Kendall do not differentiate the importance of discordant pairs across different relevance levels, while nDCG@10 assigns larger gains to documents with higher relevance labels, we calculate the consistency metrics focusing on documents with human annotations larger than 2.

As shown in Tab. \ref{tab:consis}, providing the input document list, the annotations obtained by giving explicit annotation guidelines are more consistent with human annotations and achieve better performance on the test set compared to directly letting the LLM perform ranking. Furthermore, given the annotation guidelines, compared to annotating documents sequentially for each query-document pair, providing the LLM with the complete document list and having it annotate all documents at once results in higher consistency and better ranking performance. Here, we only present the overall consistency results. The relative consistency levels of different annotation strategies across query frequencies remain consistent with the overall trends.
\begin{table}[!t]
    \centering
    \vspace{5mm}
    \caption{Comparison of consistency metrics between different LLM annotation strategies and human annotations}
    \vspace{-3mm}
    \begin{tabular}{cccc}\toprule
    \multirow{2}{*}{\makecell{Annotation\\Strategies}} & \multicolumn{2}{c}{Ranking Consistency} & Ranking Performance \\\cmidrule(r){2-3}\cmidrule(r){4-4} & Spearman & Kendall & nDCG@10 \\\midrule
    PointAnn &0.2651&0.2509&0.8481\\
    ListAnn &\textbf{0.3022}&\textbf{0.2853}&\textbf{0.8662}\\
    ListRank &0.2534&0.2211&0.8577
     \\\bottomrule
    \end{tabular}
    \label{tab:consis}
    \vspace{-6mm}
\end{table}
\section{Can LLM Annotations Replace Clicks for LTR}
In this section, we describe the training methods for click data and LLM annotations, the design of the ranking models, and three comparison dimensions for LTR: query frequency, relevance signal modeling, and relevance discrimination. Together, these analyses provide a comprehensive comparison between models trained with click data and those trained with LLM annotations, with the goal of examining whether LLM annotations can replace clicks for LTR.

\subsection{Training Methods}
In this section, we introduce the specific training methods employed for different LLM-based annotation approaches, as well as various ULTR methods for click data.

For a given query $q$ and its document list $D$, the loss function calculated for a document $d$ is denoted as \(\mathcal{L}\). We use $l_d$ and $c_d$ to represent the LLM annotation and click label of $d$, respectively. We denote the ranking model as $f$.

\noindent\textbf{LLM Annotation-Supervised Training.} For PointAnn, ListAnn, and ListRel, we adopt the listwise loss for training \cite{ai2018learning}:
\begin{equation}
    \mathcal{L}_{list}=-l_d\cdot log\frac{e^{f(q,d)}}{\sum_{d'\in D}e^{f(q,d')}}\text{.}
\end{equation}

For ListRank, we employ two types of training methods: 1) use $10-\text{rank}_d$ as the relevance label and training with the listwise loss, where $\text{rank}_d$ represents the rank of $d$ in the permutation; and 2) training with the RankNet loss \cite{burges2010ranknet}:
\begin{equation}
    \mathcal{L}_{pair}=\sum_{d_i\in D}\sum_{d_j\in D}\mathrm{I}(l_{d_i}<l_{d_j})\cdot log(1+exp(f(q,d_i)-f(q,d_j)))\text{.}
\end{equation}
RankNet loss is a pairwise loss that optimizes the relative relevance between pairs of documents. A ranking list of length $n$ yields \(\frac{n(n-1)}{2}\) document pairs.

\noindent\textbf{Unbiased Learning to Rank (ULTR).}\label{ultr} ULTR methods are mostly based on the user examination hypothesis \cite{richardson2007predicting}:
\begin{equation}
P(c_d=1)=P(r_d=1)\cdot P(o_d=1)\text{,}
\end{equation}
which states that a document $d$ is clicked ($c_d=1$) if and only if it is relevant ($r_d=1$) and observed by the user ($o_d=1$). To obtain an unbiased ranking model $f$, the key lies in accurately estimating the propensity model $g$, i.e., $P(o_d=1)$. Note that we uniformly utilize the listwise cross-entropy loss function with the inverse propensity weighting (IPW) \cite{joachims2017unbiased,wang2018position} for training:
\begin{equation}
    \mathcal{L}_{IPW}=\mathcal{L}(f(q,d);\frac{c_d}{g})\text{.}
\end{equation}

Many existing studies primarily focus on mitigating position bias, assuming that the probability of a document being observed depends solely on its position. However, real click data arises from more complex user interactions, making the biases more intricate and not limited to position bias alone. On the one hand, the observation probability of a document depends not only on its position but also on the user behavior of surrounding documents. On the other hand, due to the stronger performance of ranking models in real-world scenarios, there exists a strong coupling between position and relevance.

Therefore, we adopt the following ULTR methods:
\begin{itemize}
\item \textbf{Naive}: is a commonly used ULTR method, which directly uses clicks as relevance, without additional debiasing:
\begin{equation}
\mathcal{L}_{Naive}=\mathcal{L}(f(q,d);c_d)\text{.}
\end{equation}
\item \textbf{Dual Learning Algorithm (DLA)} \cite{ai2018unbiased}: is a commonly used method to mitigate position bias, simultaneously training an unbiased ranking model and an unbiased propensity model:
\begin{equation}
\mathcal{L}_{DLA}=\mathcal{L}(f(q,d);\frac{g_1}{g_{k_d}}c_d)+\mathcal{L}(g(k_d);\frac{f(q,d_1)}{f(q,d)}c_d)\text{,}
\end{equation}
where $k_d$ represents the position of the document $d$ in the document list.
\item \textbf{Interactional Observation-Based Model (IOBM)} \cite{chen2021adapting}: compared to DLA, IOBM considers positions and clicks of both the document itself and the surrounding documents to model the propensity model:
\begin{equation}
\mathcal{L}_{IOBM}=\mathcal{L}(f(q,d);\frac{g_1}{g_{k_d}}c_d)+\mathcal{L}(g(k_d;\mathbf{c_q});\frac{f(q,d_1)}{f(q,d)}c_d)\text{,}
\end{equation}
where $c_q$ represents the click information of the document list under query $q$.
\item \textbf{Unconfounded Propensity Estimation (UPE)} \cite{luo2023unconfounded}: leverages backdoor adjustment to address propensity overestimation:
\begin{equation}
\mathcal{L}_{UPE}=\mathcal{L}(f(q,d);\frac{g_{do(k_1)}}{g_{do(k_d)}}c_d)+\mathcal{L}(g(k_d);\frac{f(q,d_1)}{f(q,d)}c_d)\text{.}
\end{equation}
\item \textbf{Observation Dropout (Drop)} and \textbf{Gradient Reversal (GradRev)} \cite{zhang2023towards}: these two methods aim to disentangle observation and relevance. \textbf{Drop} introduces a dropout layer following the propensity model, whereas \textbf{Gradrev} appends a negative gradient reversal layer after the propensity model, where the negative gradient originates from the fitting loss $\mathcal{L}'_{rev}$ of the propensity model with respect to relevance:
\small
\begin{equation}
\begin{aligned}
&\mathcal{L}_{Drop}=\mathcal{L}(f(q,d);\frac{g_1}{g_{k_d}}c_d)+\mathcal{L}(Dropout(g(k_d));\frac{f(q,d_1)}{f(q,d)}c_d)\text{,}\\
&\mathcal{L}_{Gradrev}=\mathcal{L}(f(q,d);\frac{g_1}{g_{k_d}}c_d)+\mathcal{L}(g(k_d);\frac{f(q,d_1)}{f(q,d)}c_d)+\mathcal{L}'_{rev}\text{.}    
\end{aligned}
\end{equation}
\end{itemize}
\begin{table*}[t]
    \centering
    \vspace{4mm}
    \caption{Overview of different groups of LTR features from real click data.}
    \vspace{-3mm}
    \begin{tabularx}{\textwidth}{llX}\toprule
    Feature Group & Feature Name & Feature Description\\\midrule
    \multirow[c]{9}{*}{Q-D Matching}&TF, IDF, TF·IDF&The sum of term frequency (TF), inverse document frequency (IDF), and TF·IDF of query terms in title, content, and the whole document.\\
    & BM25 \cite{robertson1994some} & The scores of BM25 on title, content and the whole document. \\
    & LMJM, LMDIR, LMABS \cite{zhai2004study}& The scores of language model (LM) with different smoothing methods—Jelinek-Mercer (JM), Dirichlet (DIR), and absolute discounting (ABS)—on title, content, and the whole document. \\
    & proximity1 \cite{chen2023thuir} & Averaged proximity score and positions of query terms within the whole document, computed both including and excluding stop words. \\
    & proximity2 \cite{chen2023thuir} & Number of query bigrams appearing in the whole document, computed both including and excluding stop words, within a window of 5 and 10 words. \\\midrule
    \multirow[c]{3}{*}{Document}&Length&The length of title, content, URL, and the whole document.\\
    &Slash&The number of slashes in the URL.\\
    &PageRank \cite{page1999pagerank}&The PageRank score of the URL domain.\\\bottomrule
    \end{tabularx}
    \label{tab:feat}
\end{table*}
\subsection{Ranking Models}
State-of-the-art ranking models commonly adopt a cross-encoder architecture \cite{alaparthi2020bidirectional}, where the input is concatenated as ``[CLS] query [SEP] document [SEP]''. The [CLS] token representation is extracted and used to predict the relevance between the query and the document. We choose ERNIE-3.0 Base \cite{sun2021ernie} as the backbone due to its strong semantic representation capabilities for Chinese text.

To investigate the differences in relevance signal modeling capabilities of models trained with different types of annotations, we introduce LTR features at different levels as diagnostic signals and incorporate them by concatenating with the relevance score extracted from a fine-tuned cross-encoder model. Although tree-based models \cite{burges2010ranknet} are commonly more effective for learning-to-rank with handcrafted LTR features, we adopt a lightweight Deep Neural Network (DNN)-based ranking model for click-based supervision, as existing debiasing methods for real click data are less effective when applied to tree-based approaches \cite{hu2019unbiased,zou2022large}. For a fair comparison, we also adopt a DNN-based ranker for models trained with LLM annotations, using the same combined representations.

\subsection{Comparison Dimensions for LTR}\label{sec:comp}

To examine whether LLM annotations can replace user clicks for LTR, we compare the two supervision sources from three perspectives:

\textbf{Query Frequency.}
Click data and LLM annotations may be useful for different query types. Clicks reflect not only query-document semantic matching, but also document-level factors such as authority, content quality, and user-perceived utility. In contrast, LLMs are strong at semantic relevance assessment but may have limited access to such document-level signals, e.g., authority inferred from URL domains.
This difference may be especially important across query frequencies. For high-frequency queries, search engines often return many relevant results, making document-level signals important for distinguishing among them. Clicks may therefore provide stronger supervision in such cases. For middle- and low-frequency queries, however, click signals are sparse, making LLM annotations potentially more useful. We thus compare the two supervision sources across different query-frequency groups.

\textbf{Relevance Signal Modeling.}
Click data and LLM annotations may capture different relevance signals. To analyze this, we introduce query-document (Q-D) matching features, such as BM25 \cite{robertson1994some}, and document-level features, such as PageRank \cite{page1999pagerank}, to complement the query-document matching signal with semantic embeddings. By comparing the performance gains from these feature groups, we examine whether models trained with different supervision sources better capture semantic matching or document-level relevance signals. 

\textbf{Relevance Discrimination.}
Click data is usually collected from top-ranked results examined by users, often the top 10 \cite{zou2022large}. Consequently, click-based training data contains many relevant or partially relevant documents but few truly irrelevant ones. Models trained on such data may learn fine-grained preferences among relevant documents, but their ability to distinguish relevant from non-relevant documents remains unclear.
To study this, we introduce true negative samples into the TianGong-ST test set. These samples are documents that appear in search logs but were never clicked for any query and were manually verified as irrelevant. We select 10 representative true negatives from the top-ranked results that receive no clicks across multiple query sessions. Although the sample size is limited due to manual verification cost, the observed patterns are consistent across these examples, providing preliminary evidence about the relevance discrimination ability of different training approaches. We also examine how adding such true negatives during training affects ranking performance.
\section{Experimental Setup}
\noindent\textbf{Dataset.} We conduct experiments on two Chinese search-engine click datasets: the public TianGong-ST \cite{chen2019tiangong} and a private Baidu-Click dataset. Notably, we focus on Chinese datasets due to the scarcity of publicly available click data in other languages. TianGong-ST contains 18 days of Sogou search logs, including the top-10 results and click feedback, with 2k queries annotated on a 0–4 relevance scale. After filtering, about 260k queries with at least one click are used for training. Baidu-Click is collected from Baidu with a similar setup: roughly 260k queries from one week for training, and around 7k queries from the following week with top-50 ranking-stage results annotated on the same relevance scale. Human annotations in both datasets are split into validation and test sets with a 3:7 ratio. For query-frequency analysis, Baidu-Click uses predefined bins, while TianGong-ST applies equal-frequency binning to partition queries into the same three coarse-grained categories (high, middle, low). Representative LTR features for both datasets are extracted as summarized in Tab. \ref{tab:feat}. More comprehensive feature sets are left for future investigations.

\noindent\textbf{Metrics.} To evaluate ranking performance, we report nDCG \cite{jarvelin2002cumulated} at positions 1, 3, 5, and 10. We further report these metrics separately for queries of different frequency categories. All experimental results are averaged over three random seeds, and we report the mean values. Due to space constraints, we do not report confidence intervals for all results, but we note that the relative performance trends remain consistent across runs.

\noindent\textbf{Implementation Details.} Since the document content in TianGong-ST is crawled from the original web page, it contains substantial irrelevant information (e.g., navigation bars) unrelated to the main content. To address this, we employ DeepSeek-R1-Distill-Qwen-32B to extract the core content, effectively generating a document summary. When fine-tuning ERNIE-3.0 Base, we set the maximum sequence length to 160 for TianGong-ST and 640 for Baidu-Click, respectively.

The [CLS] representation extracted from the ERNIE-3.0 Base is fed into a three-hidden-layer DNN with hidden sizes of [512, 256, 128] to produce the final relevance score. For the ranking model that takes the concatenation of LTR features and the relevance score extracted from the fine-tuned ERNIE-3.0 Base as input, we adopt different model architectures under different supervision. For click data, the concatenated features are first mapped to a 64-dimensional vector and then passed through a three-hidden-layer DNN with hidden sizes of [32, 16, 8] to generate the final relevance score. Each hidden layer is followed by an ELU activation function. Clicks or LLM annotations are then used to train the neural rankers for comparison. 

We implement the ULTR methods described in Sec. \ref{ultr} using the ULTRA toolkit \cite{tran2021ultra}. For each ULTR method, we keep all model-specific parameters consistent with their original papers. We set the batch size to 64, use the AdamW optimizer with a learning rate of 3e-5, and apply early stopping if no improvement is observed on the validation set for 3 consecutive epochs. The maximum number of training epochs is set to 8.
\section{Results and Analysis}
We compare models trained solely on click data and those trained solely on annotations according to the analytical dimensions proposed in Section \ref{sec:comp}. The experimental results reveal systematic differences in ranking performance, relevance signal modeling, and the ability to distinguish true negative samples, thereby motivating subsequent joint training.
\begin{table*}[!t]
\setlength{\tabcolsep}{1.0pt}
\small
    \centering
    \vspace{4mm}
    \caption{Performance comparison of various methods trained with click data and LLM annotations. \textbf{Bold} and \underline{underline} indicate the best and second-best methods within the same type of annotation; for the Gradrev method, ``$*$'' indicates significantly better performance than the best model trained with LLM annotations ($p \leq 0.05$). Results are averaged over three random seeds.}
    \vspace{-3mm}
    \begin{tabular}{cccccccccccccccccc}\toprule
    \multirow{2}{*}{Datasets}&\multirow{2}{*}{Methods}&\multicolumn{4}{c}{nDCG@1}&\multicolumn{4}{c}{nDCG@3}&\multicolumn{4}{c}{nDCG@5}&\multicolumn{4}{c}{nDCG@10}\\\cmidrule(r){3-6}\cmidrule(r){7-10}\cmidrule(r){11-14}\cmidrule(r){15-18}&&All&High&Mid&Low&All&High&Mid&Low&All&High&Mid&Low&All&High&Mid&Low\\\midrule
    \multirow{11}{*}{TianGong-ST}
    &Naive& 0.6992&0.9224&0.6849&0.6296&0.7098&0.8659&0.6930&0.6680&0.7389&0.8645&{0.7229}&0.7078&\underline{0.8673}&0.9433&\textbf{0.8597}&0.8452\\
    &DLA&0.6990&0.9224&0.6746&0.6395&0.7068&{0.8785}&0.6823&0.6669&0.7367&0.8613&0.7184&0.7082&0.8660&0.9437&0.8567&0.8447 \\
    &IOBM&0.6938&0.9224&0.6596&\textbf{0.6423}&0.7042&0.8753&0.6670&\textbf{0.6773}&0.7359&{0.8646}&0.7129&\textbf{0.7106}&0.8652&0.9429&0.8504&\textbf{0.8480} \\
    &UPE&\textbf{0.7246}&\underline{0.9230}& \textbf{0.7330}& \underline{0.6412}&\underline{0.7134}&\underline{0.8884}&\textbf{0.6963}&0.6648&0.7391&0.8664& \textbf{0.7313}& 0.6991&\underline{0.8664}& \underline{0.9459}& 0.8567& 0.8445 \\
    &Drop&\underline{0.7067}&0.9224&\underline{0.6969}&0.6352&{0.7116}&0.8742&{0.6944}&0.6677&\textbf{0.7408}&0.8621&\underline{0.7269}&0.7092&0.8668&0.9425&0.8587&0.8452 \\
    &Gradrev&{0.7013}&$\textbf{0.9232}^*$&{0.6789}&${0.6403}^*$&\textbf{0.7150}&$\textbf{0.8926}^*$&\underline{0.6949}&\underline{0.6683}&\underline{0.7399}&\textbf{0.8710}&0.7210&\underline{0.7097}&\textbf{0.8684}&$\textbf{0.9469}^*$&\underline{0.8590}&\underline{0.8467} \\\cmidrule{2-18}
    &PointAnn&0.6830&0.6950&0.7393&0.6214&0.7234&0.7474&0.7573&0.6801&0.7615&0.7865&0.7824&0.7309&0.8643&0.8690&0.8806&0.8502\\
    &ListAnn&\textbf{0.7165}&\textbf{0.8332}&\textbf{0.7782}&0.6101&\textbf{0.7570}&\textbf{0.8690}&\underline{0.7802}&\underline{0.6913}&\textbf{0.7869}&\textbf{0.8837}&\underline{0.7999}&0.7373&\textbf{0.8817}&\textbf{0.9453}&\textbf{0.8886}&0.8531 \\
    &ListRank\_list&0.6613&\underline{0.5227}&0.7459&\underline{0.6278}&0.7407&\underline{0.7559}&0.7780&0.6972&0.7675&\underline{0.7665}&0.7941&\underline{0.7409}&0.8632&\underline{0.8553}&0.8783&\underline{0.8546} \\
    &ListRank\_pair&\underline{0.6704}&0.4996&\underline{0.7657}&\textbf{0.6382}&\underline{0.7447}&0.7471&\textbf{0.7873}&\textbf{0.7006}&\underline{0.7715}&0.7604&\textbf{0.8020}&\textbf{0.7448}&\underline{0.8664}&0.8508&\underline{0.8851}&\textbf{0.8578} \\
    &ListSel&0.5820&0.4561&0.6470&0.5635&0.6319&0.5575&0.6567&0.6347&0.6993&0.6788&0.7173&0.6889&0.8179&0.7849&0.8231&0.8261 \\\cmidrule{2-18}
    &Zero-shot&0.6442&0.5833&0.7036&0.6070&0.7002&0.7117&0.7235&0.6723&0.7455&0.7517&0.7606&0.7278&0.8579&0.8464&0.8714&0.8486 \\\midrule
    \multirow{2}{*}{Baidu-Click}
    &Gradrev&0.5460&$0.5465^*$&0.5749&0.5276&0.5873&$0.5831^*$&0.6148&0.5715&0.5920&$0.5954^*$&0.6253&0.5675&0.6057&$0.6270^*$&0.6383&0.5714 \\\cmidrule{2-18}
    &ListAnn&0.5781&0.5401&0.5946&0.5849&0.6147&0.5779&0.6330&0.6202&0.6171&0.5921&0.6395&0.6147&0.6311&0.6201&0.6540&0.6214 \\\bottomrule
    \end{tabular}
    \label{tab:performance}
    \vspace{-3mm}
\end{table*}
\subsection{Performance Comparison}
Tab. \ref{tab:performance} presents the overall and query-frequency-specific performance of various ULTR methods and LLM annotation-based training approaches. We first separately compare the training approaches within each annotation type. When trained on real-world click data, the performance gap between DLA and the Naive baseline is minimal, indicating that addressing only position bias is insufficient for real-world click data. In contrast, UPE, Drop, and GradRev consistently maintain performance gains over the Naive method, suggesting that mitigating the coupling between relevance and position induced by strong ranking models is more critical for real-world click data. Overall, the GradRev method achieves the best performance.

When trained on LLM annotations, models trained with multi-level annotations (PointAnn, ListAnn, and ListRank) outperform those trained with single-level annotations (ListSel). Among multi-level methods, PointAnn yields the worst performance, highlighting the advantage of listwise input strategies. Meanwhile, the two training approaches based on the ListRank annotation strategy perform similarly. Still, they are inferior to ListAnn, possibly due to overemphasizing the individual document relevance when the generated ranking quality is suboptimal. The model performance across different annotation strategies with LLMs largely aligns with their agreement with human annotations in Tab. \ref{tab:consis}, indicating that the consistency with human annotations can serve as a reliable criterion for selecting annotation strategies. 

\enlargethispage{2\baselineskip}
We then compare the performance of models trained on click data and LLM annotations. Overall, models supervised by LLM annotations outperform those supervised by clicks. However, considering query frequencies, LLM annotation-supervised models significantly excel on middle- and low-frequency queries, while click-supervised models retain an advantage on high-frequency queries. This suggests LLM annotations cannot yet fully replace clicks. The fundamental reason for this performance difference derives from the differences in their abilities to capture relevance. In real-world scenarios, ranking models generally perform better on high-frequency queries, returning more relevant results, thus document-level signals like authority and quality become crucial for distinguishing relevance. Click data can capture these signals, whereas LLM annotations struggle to do so. Attempts to incorporate numeric signals like PageRank into LLM annotations were unsuccessful, likely due to LLMs’ limited ability to interpret numerical data. Conversely, middle- and low-frequency queries rely more on semantic matching, highlighting the strength of LLM annotations in capturing semantic relevance.

In subsequent experiments, we consistently use the GradRev method on click data and the ListAnn annotation strategy for training. We validate the performance differences between the two on the Baidu-Click dataset. As shown in the Tab. \ref{tab:performance}, models trained with click data maintain their advantage on high-frequency queries, while models trained with LLM annotations perform better on middle- and low-frequency queries, which is consistent with the conclusions above.
\begin{figure}[!t]
    \centering
    \begin{subfigure}[b]{0.23\textwidth}
      \includegraphics[width=\textwidth]{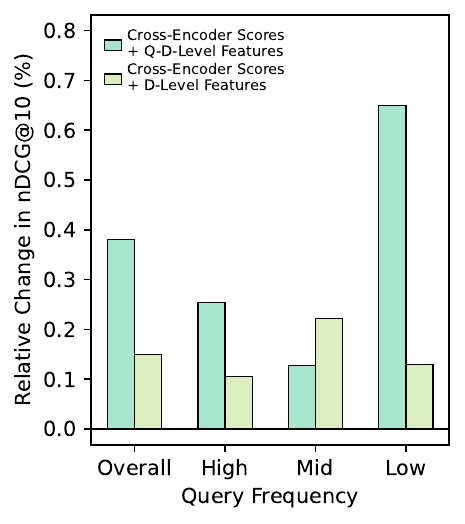}
      \vspace{-7mm}
      \caption{Click Data}
      \label{fig:feat_click}
    \end{subfigure}%
    ~
    \begin{subfigure}[b]{0.23\textwidth}
      \includegraphics[width=\textwidth]{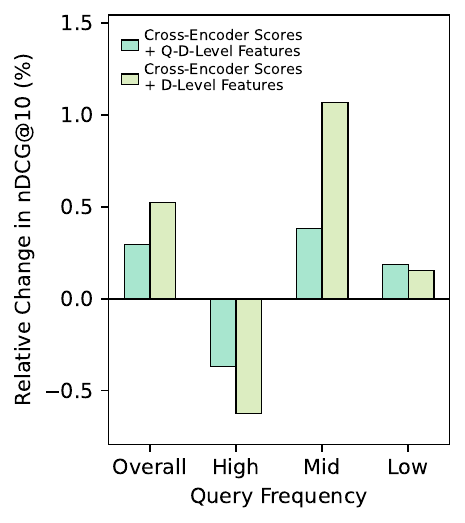}
      \vspace{-7mm}
      \caption{LLM Annotations}
      \label{fig:feat_llm}
    \end{subfigure}
    \vspace{2mm}
    \caption{Effects of document-level and query-document-level LTR features on models on the TianGong-ST dataset.}
    \label{fig:feat}
\end{figure}
\subsection{Impact of LTR Feature Groups}
After extracting the semantic relevance scores from the fine-tuned cross-encoders, we combine them with lexical matching features and document-level features, respectively, and compare the relative performance gains over the original models, as shown in Fig. \ref{fig:feat}.

\enlargethispage{1\baselineskip}
For click-based models, incorporating lexical matching features yields larger performance improvements, with the most pronounced gains observed on low-frequency queries. This suggests that models fine-tuned on click data may already possess a stronger capability to differentiate relevance from document-level signals, while their ability to capture term-based matching remains limited and can be further enhanced. 

In contrast, models fine-tuned on LLM annotations already exhibit strong capability in distinguishing relevance from semantic matching signals, making the additional gains from lexical matching features less evident compared to the click-based setting. Introducing document-level features, however, leads to performance improvements on middle-frequency queries, possibly because LLM annotations implicitly encode coarse-grained document-level distinctions. The performance degradation observed on high-frequency queries may indicate that such document-level signals are not sufficiently robust, and that incorporating more fine-grained document-level features can conflict with the supervision provided by LLM annotations.
\begin{figure}[!t]
    \centering
    \begin{subfigure}[b]{0.36\textwidth}
      \includegraphics[width=\textwidth]{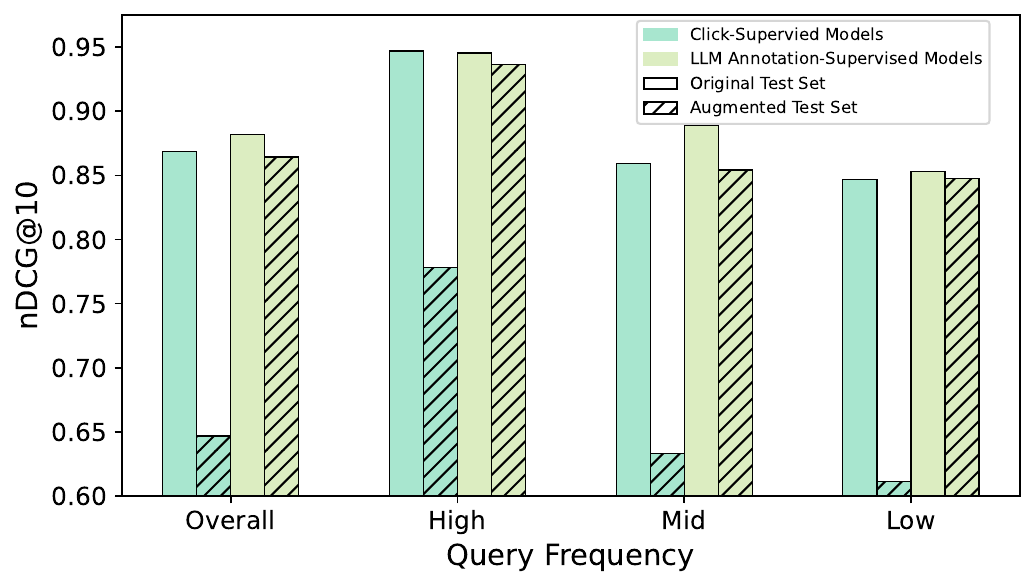}
      \vspace{-7mm}
      \caption{Incorporating True Negatives in the Test Set}
      \label{fig:neg_test}
    \end{subfigure}%
    \vspace{6mm}
    \\
    ~
    \begin{subfigure}[b]{0.36\textwidth}
      \includegraphics[width=\textwidth]{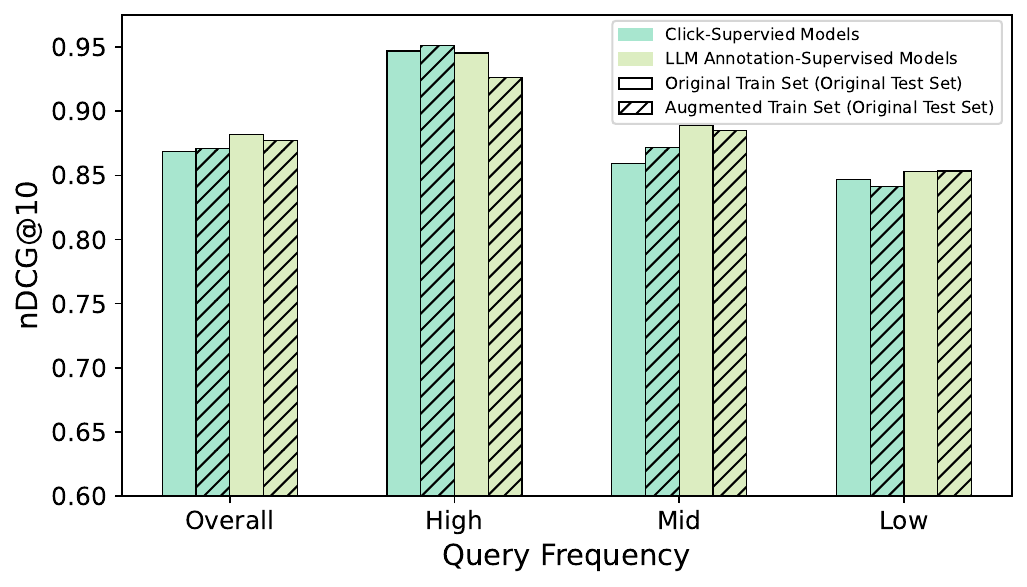}
      \vspace{-7mm}
      \caption{Incorporating True Negatives in the Train Set}
      \label{fig:neg_train}
    \end{subfigure}
    \vspace{2mm}
    \caption{Effects of true negatives in training and test sets on model performance on the TianGong-ST dataset.}
    \label{fig:true_neg}
\end{figure}
\subsection{Exploring Relevance Discrimination via True Negatives}
As shown in Fig. \ref{fig:true_neg}(\subref{fig:neg_test}), after adding true negatives to the test set, the performance of models trained with both annotation types declines. However, the performance drop of click-supervised models is significantly larger than that of LLM annotation-supervised models. This indicates that models trained with LLM annotations not only distinguish relevance differences within relevant results but also effectively separate relevant from non-relevant documents, whereas click-supervised models have weaker discrimination ability between relevant and non-relevant documents.

Furthermore, we introduce 10 in-batch negatives into the training set to enhance the models’ ability to differentiate relevant from non-relevant documents and explore the impact on both types of trained models. As illustrated in Fig. \ref{fig:true_neg}(\subref{fig:neg_train}), for click-supervised models, incorporating true negatives in training further improved ranking performance by strengthening relevance discrimination. In contrast, since LLM annotation-supervised models already possess this ability, adding true negatives caused the models to focus more on the obvious boundary between relevant and non-relevant documents, which in turn led to an overall performance decline, as the models require more nuanced discrimination within relevant documents.
\section{Hybrid Training: Integrating Clicks and LLM Annotations}
The above experimental results indicate that under the cross-encoder architecture, the relevance discrimination capabilities of models trained on click data and those trained on LLM annotations align closely with the inherent strengths of their respective supervision signals. This alignment manifests as performance advantages on queries of different frequencies, highlighting a natural complementarity between the two types of annotation. These findings suggest that queries with varying frequencies may benefit from distinct supervision signals. 

Motivated by this, we further investigate effective strategies to integrate click data and LLM annotations within the cross-encoder framework, aiming to train ranking models that possess strong capabilities in capturing both semantic matching signals and document-level information simultaneously. In this section, we provide a detailed description and experimental results of the two integration approaches we propose: Annotation-aware Data Scheduling (ADS) and Frequency-Aware Multi-Objective Learning (FAMOL). 
\begin{figure}[!t]
    \centering
    \vspace{-2mm}
    \includegraphics[width=0.9\linewidth]{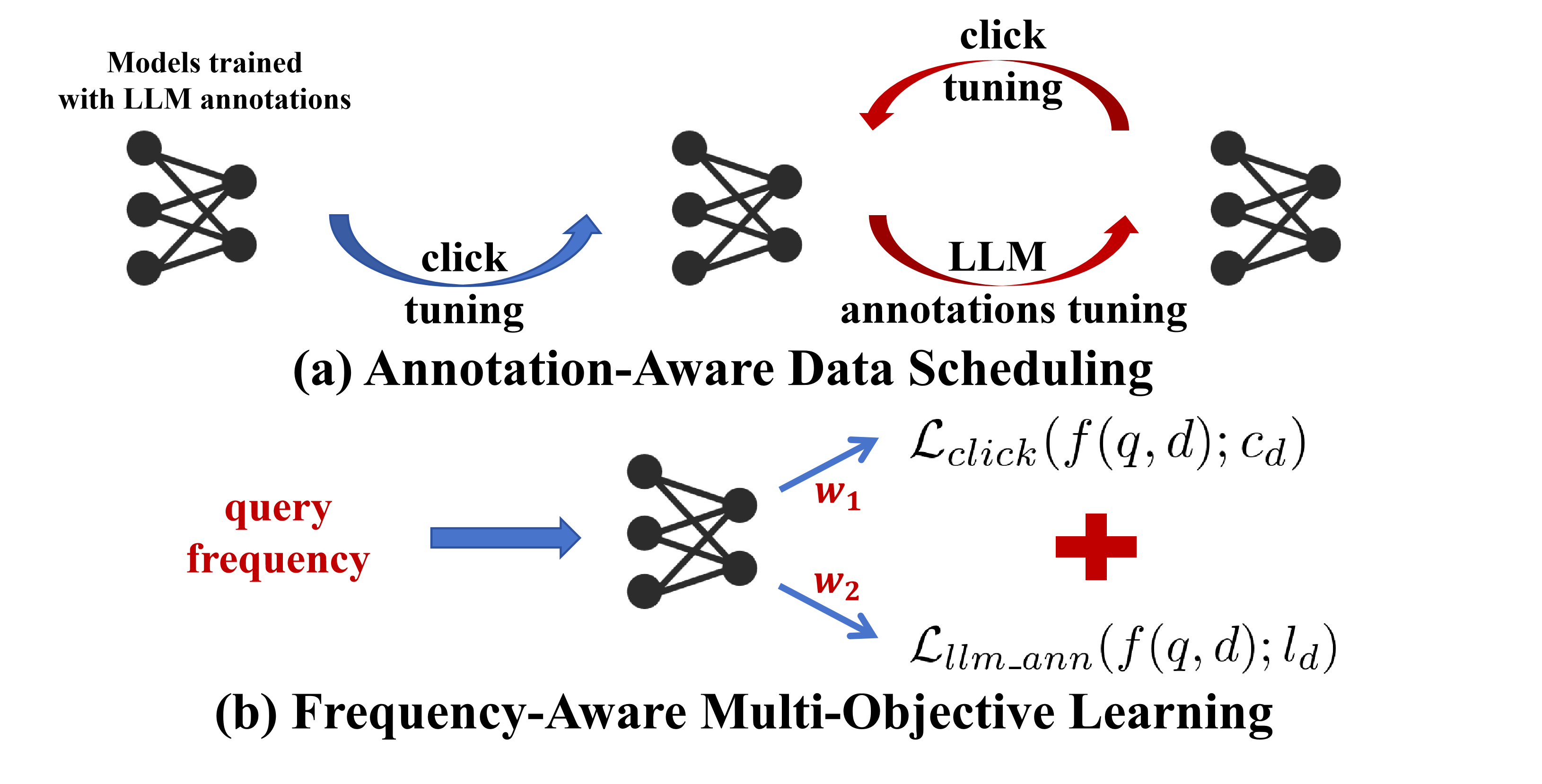}
    \vspace{-5mm}
    \caption{Hybrid training methods: (a) Annotation-aware data scheduling and (b) Frequency-aware multi-objective learning.}
    \label{fig:hybrid}
    \vspace{1.2mm}
\end{figure}
\subsection{Annotation-Aware Data Scheduling}
Motivated by the following two considerations: 1) LLM annotations exhibit strong capability in capturing semantic matching between queries and documents, whereas click data captures both semantic matching and document-level signals such as authority; and 2) although obtaining LLM annotations via even open-source LLMs is still relatively more expensive compared to click data, their annotation quality is superior.

We propose the Annotation-Aware Data Scheduling (ADS) method and adopt the following specific training procedure: (1) First, we warm-start the ranking model by training on 120k LLM annotations (all queries) for 3 epochs with batch size 64, establishing a solid foundation in semantic matching-based relevance judgment; (2) Then, to further enhance the model’s ability to discriminate relevance by incorporating document-level signals while mitigating catastrophic forgetting, we perform alternating training: each alternation consists of 1 epoch on 10k high-frequency click data followed by 1 epoch on 120k middle- and low-frequency LLM annotations; (3) We perform 2-3 alternation cycles, with early stopping applied based on validation set performance. The learning rate is set to 3e-5 with AdamW optimizer, following the same hyperparameters as single-supervision training.

\begin{figure}[!t]
    \centering
    \begin{subfigure}[b]{0.25\textwidth}
      \includegraphics[width=\textwidth]{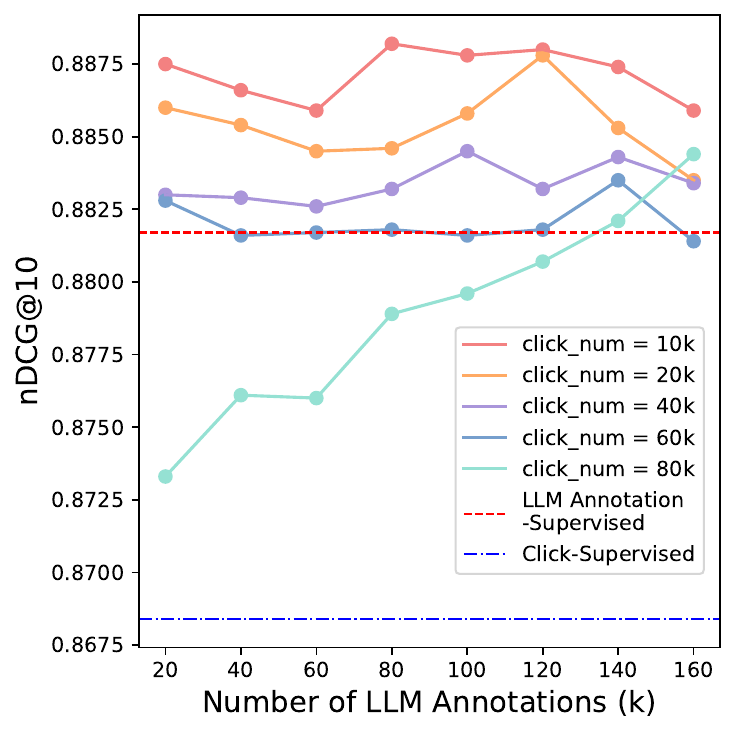}
      \vspace{-6mm}
      \caption{Performance with Varying Amounts of Click Data and LLM Annotations}
      \label{fig:schedule_total}
    \end{subfigure}%
    ~
    \begin{subfigure}[b]{0.22\textwidth}
      \begin{subfigure}[b]{1.0\textwidth}
          \includegraphics[width=\textwidth]{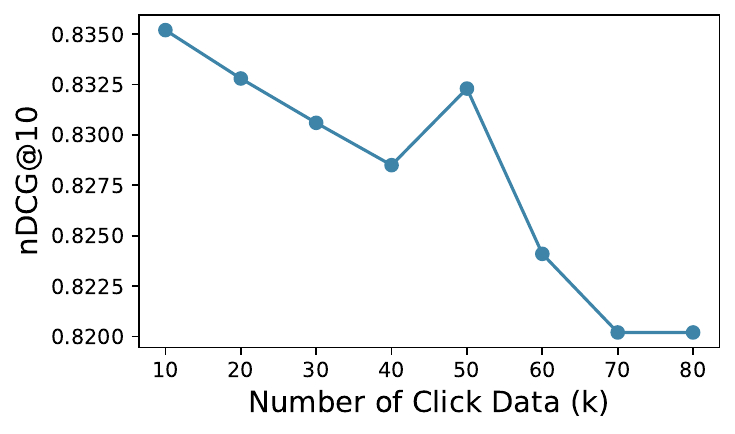}
          \vspace{-6mm}
          \captionsetup{font=footnotesize}
          \caption{Performance Using High-Frequency Click Data}
          \label{fig:schedule_click}
      \end{subfigure}%
      \vspace{6mm}
        \\
        ~
      \begin{subfigure}[b]{1.0\textwidth}
          \includegraphics[width=\textwidth]{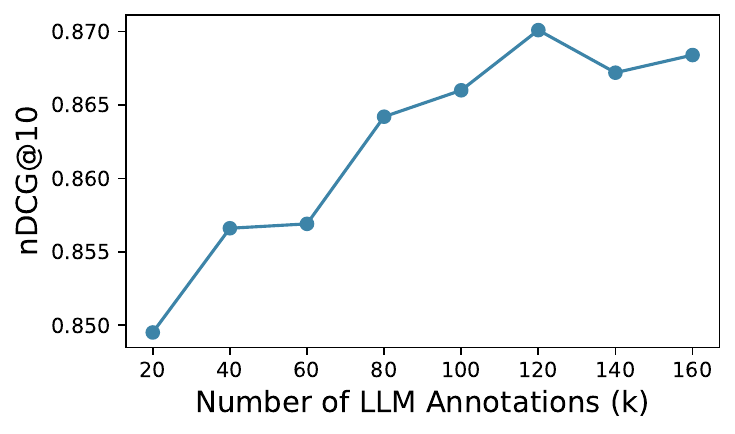}
          \vspace{-6mm}
          \captionsetup{font=footnotesize}
          \caption{Performance Using Middle- and Low-Frequency LLM Annotations}
          \label{fig:schedule_llm}
      \end{subfigure}
    \end{subfigure}
    \vspace{4mm}
    \caption{Impact of Varying Volumes of Click and LLM Annotations on Annotation-Aware Data Scheduling on the TianGong-ST dataset..}
    \label{fig:schedule}
\end{figure}

We compare fine-grained mixing, which combines click data and LLM annotations within each batch, with coarse-grained alternation, which switches periodically between the two annotation types. Since fine-grained mixing brings no improvement, possibly due to conflicting objectives, we adopt coarse-grained alternation and tune the click-to-LLM data ratio. Preliminary experiments in Fig. \ref{fig:schedule}(\subref{fig:schedule_total}) identify 10k click data and 120k LLM annotations as optimal. These scales are consistent with the optima in single-supervision training, as shown in Fig. \ref{fig:schedule}(\subref{fig:schedule_click}), suggesting that optimal data scheduling requires each supervision signal to use a data scale similar to that in its corresponding single-supervision setting.
\subsection{Frequency-Aware Multi-Objective Learning} 
Queries of different frequencies require different types of supervision signals. Based on this, we design a frequency-aware weighting mechanism formalized as Eq. \ref{eq:hybrid} and shown in Fig. \ref{fig:hybrid}(b).
\begin{equation}
\label{eq:hybrid}
\begin{aligned}
&w_1,w_2=sigmoid(\text{FFN}(\text{Emb}(q_{freq})))\text{,}
\\
&\mathcal{L}=w_1\cdot\mathcal{L}_{click}(f(q,d);c_d)+w_2\cdot\mathcal{L}_{llm\_ann}(f(q,d);l_d)\text{.}
\end{aligned}
\end{equation}
Let $q_{freq}$ denote the query frequency. We set the embedding dimension to 16, and FFN is a two-layer DNN with sizes [64, 2].
\begin{table}[!t]
    \setlength{\tabcolsep}{0.2pt}
    \small
    \centering
    \vspace{6mm}
    \caption{Performance comparison (nDCG@10) of methods combining click data and LLM annotations. \textbf{Bold} indicates the best-performing method within the same supervision type. ``$*$'' indicates that one single-supervision signal significantly outperforms the other under separate training, while ``$\dag$'' indicates that a hybrid training method significantly outperforms the best single-supervision method ($p \leq 0.05$). Results are averaged over three random seeds.}
    \vspace{-3mm}
    \begin{tabular}{ccccccccc}\toprule
    \multirow{2}{*}{Methods}&\multicolumn{4}{c}{TianGong-ST}&\multicolumn{4}{c}{Baidu-Click}\\
    \cmidrule(r){2-5}\cmidrule(r){6-9}&All&High&Mid&Low&All&High&Mid&Low\\\midrule
    \multicolumn{2}{c}{\textit{Single Supervision}}&&&&&&\\
    Gradrev&0.8684&$\textbf{0.9469}^*$&0.8590&0.8467& 0.6057&$\textbf{0.6270}^*$&0.6383&0.5714 \\
    ListAnn&$\textbf{0.8817}^*$&0.9453&$\textbf{0.8886}^*$&$\textbf{0.8531}^*$& $\textbf{0.6311}^*$&0.6201&$\textbf{0.6540}^*$&$\textbf{0.6214}^*$ \\\midrule
    \multicolumn{2}{c}{\textit{Hybrid Supervision}}&&&&&&\\
    ADS&${0.8878}^{\dag}$&$\textbf{0.9592}^{\dag}$&$\textbf{0.8952}^{\dag}$&${0.8560}^{\dag}$&- &- &- &- \\
    FAMOL&$\textbf{0.8897}^{\dag}$&${0.9533}^{\dag}$&${0.8917}^{\dag}$&$\textbf{0.8649}^{\dag}$&$0.6333^{\dag}$&0.6275&$0.6589^{\dag}$&0.6200 \\\bottomrule
    \end{tabular}
    \label{tab:hybrid}
\end{table}

We train for 8 epochs with batch size 64 and learning rate 3e-5 using AdamW optimizer, with early stopping applied if no performance improvement is observed for 3 consecutive epochs. The learned weights at the optimal performance show only minor variation across query frequencies, with $w_2 \approx 0.97$ for high-frequency queries and $w_2 \approx 0.99$ for middle- and low-frequency queries. Despite this, the proposed fusion strategy consistently outperforms models trained with either click data or LLM annotations alone, indicating that the gains mainly come from the complementary strengths of the two supervision signals rather than explicit frequency-dependent reweighting.

\subsection{Results}
As shown in Tab. \ref{tab:hybrid}, compared to models trained solely on a single supervision signal, the frequency-aware weighting mechanism adaptively adjusts the training weights of the two supervision signals based on the current query frequency, thereby leveraging the advantages of both. Both ADS and FAMOL methods improve performance across queries of all frequency levels, indicating that both approaches can effectively leverage the strengths of the two types of annotations simultaneously. Among the two approaches, FAMOL demonstrates more consistent performance improvements. Moreover, we validate the effectiveness of FAMOL on the Baidu-Click dataset.
\section{Conclusion and Future Work}
This paper investigates whether LLM annotations can replace click data for LTR. Experiments show that click-supervised models perform better on high-frequency queries, while LLM annotation-supervised models are more effective on middle- and low-frequency queries. Further analysis reveals that LLM annotations better capture fine-grained relevance differences and distinguish relevant from irrelevant results, whereas click data benefits more from lexical features and LLM annotations may benefit from document-level features. Finally, we explore hybrid training strategies—annotation-aware data scheduling and frequency-aware multi-objective learning—which effectively combine the strengths of both annotations, with the latter showing superior performance. Future work includes developing more principled integration mechanisms and improving LLMs’ ability to incorporate document-level signals such as authority, credibility, and quality in relevance annotation and ranking.

\section*{Acknowledgments}
This work was funded by the National Natural Science Foundation of China (NSFC) under Grants No. 62302486 and No. 62441229, the Innovation Project of ICT CAS under Grants No. E361140 and E561010, and the Strategic Priority Research Program of the CAS under Grants No. XDB0680102.

\newpage
\section*{GenAI Usage Disclosure}
This work used ChatGPT by OpenAI to improve academic writing, including making expressions more natural and improving clarity and logical organization. In addition, ChatGPT was occasionally used to help identify and debug implementation issues in the code.

\bibliographystyle{ACM-Reference-Format}
\bibliography{ref}
\end{document}